\documentclass[aps,a4paper,superscriptaddress,preprint]{revtex4}
\usepackage{amsfonts,amssymb,amsmath}
\usepackage{array,dcolumn}
\usepackage{graphicx}

\usepackage{ifpdf}

\ifpdf
  \usepackage{aeguill}
  \usepackage[pdftex]{color,rotating}
  \usepackage[pdftex]{pdflscape}
  \usepackage[pdftex,backref=page,hyperfootnotes=false,colorlinks=false]{hyperref}
  \hypersetup{
    colorlinks=false,
    pdfpagemode=UseThumbs,
    plainpages=false
   }
\else
  \usepackage[dvips]{color,rotating}
  \usepackage[dvips]{lscape}
  \usepackage[dvips,backref=page,hyperfootnotes=false,colorlinks=false]{hyperref}
  \hypersetup{
    colorlinks=false,
		pdfpagemode=UseThumbs,
    plainpages=false
   }
\fi

\begin{document}

\title{A Schottky top-gated two-dimensional electron system in a nuclear spin free Si/SiGe heterostructure}

\author{J. \surname{Sailer}}
\author{V. \surname{Lang}}
\author{G. \surname{Abstreiter}}
\affiliation{Walter Schottky Institut, Technische Universit\"{a}t M\"{u}nchen, 85748 Garching, Germany}
\author{G. \surname{Tsuchiya}}
\author{K. M. \surname{Itoh}}
\affiliation{Department of Applied Physics and Physico-Informatics, Keio University 3-14-1, Hiyoshi, Kohoku-ku, Yokohama 223-8522, Japan}
\author{J. W. \surname{Ager} III}
\affiliation{Lawrence Berkeley National Laboratory, Materials Sciences Division, Berkeley, CA 94720-8197, USA}
\author{E. E. \surname{Haller}}
\affiliation{Lawrence Berkeley National Laboratory, Materials Sciences Division, Berkeley, CA 94720-8197, USA}
\affiliation{Department of Materials Science and Engineering, University of California at Berkeley, Berkeley, CA 94720-1760, USA}
\author{D. \surname{Kupidura}}
\author{D. \surname{Harbusch}}
\author{S. \surname{Ludwig}}
\affiliation{Fakult\"{a}t f\"{u}r Physik and Center for NanoScience, Ludwig-Maximilians-Universit\"{a}t M\"{u}nchen, Geschwister-Scholl-Platz 1, 80539 M\"{u}nchen, Germany}
\author{D. \surname{Bougeard}}\email[Corresponding author:]{bougeard@wsi.tum.de}
\affiliation{Walter Schottky Institut, Technische Universit\"{a}t M\"{u}nchen, 85748 Garching, Germany}

\begin{abstract}
We report on the realization and top-gating of a two-dimensional electron system in a nuclear spin free environment using $\mathrm{^{28}Si}$ and $\mathrm{^{70}Ge}$ source material in molecular beam epitaxy. Electron spin decoherence is expected to be minimized in nuclear spin-free materials, making them promising hosts for solid-state based quantum information processing devices. The two-dimensional electron system exhibits a mobility of $\mathrm{18000~cm^2/Vs}$ at a sheet carrier density of $\mathrm{4.6 \cdot 10^{11}~cm^{-2}}$ at low temperatures. Feasibility of reliable gating is demonstrated by transport through split-gate structures realized with palladium Schottky top-gates which effectively control the two-dimensional electron system underneath. Our work forms the basis for the realization of an electrostatically defined quantum dot in a nuclear spin free environment.
\end{abstract}

\date{\today}

\maketitle

Quantum dots offer a promising two-level system for applications in solid state based quantum information processing \cite{Los1998}. Within these three dimensionally confining structures, electrostatically defined quantum dots are a well studied system \cite{Han2007}, mostly in III-V-materials. A major source of decoherence in such devices is the interaction of the confined electron spin with the surrounding semiconductor host matrix, in particular with the nuclear spin bath \cite{Pet2005}. Recently, single electron devices have been reported in materials systems like Si-Ge \cite{Sim2007,Sha2008a} or C \cite{Jor2008} which contain a reduced amount of nuclear spins in their natural isotopic composition. As a next step, isotopical purification of the group-IV source materials Si, Ge and C can give access to virtually nuclear spin free materials. In this letter, we report on the realization of two-dimensional electron systems (2DES) in a nuclear spin free environment. A 2DES forms in a strained $\mathrm{^{28}Si}$ layer embedded into $\mathrm{{^{28}Si^{70}Ge}}$. The ability to control the 2DES via top-gates is demonstrated by the implementation of split-gate structures which are able to locally deplete the 2DES. Suitable voltages allow the complete pinch-off of the narrow conducting channel.

Samples are fabricated in solid source molecular beam epitaxy (MBE). The base pressure of the Riber Siva 45 MBE chamber is $\mathrm{1 \cdot 10^{-11}~mbar}$. $\mathrm{^{28}Si}$ and $\mathrm{^{70}Ge}$ are evaporated from a custom made {MBE-Komponenten} electron beam source and effusion cell respectively. 

\begin{figure}[t]
\includegraphics[width=1\columnwidth]{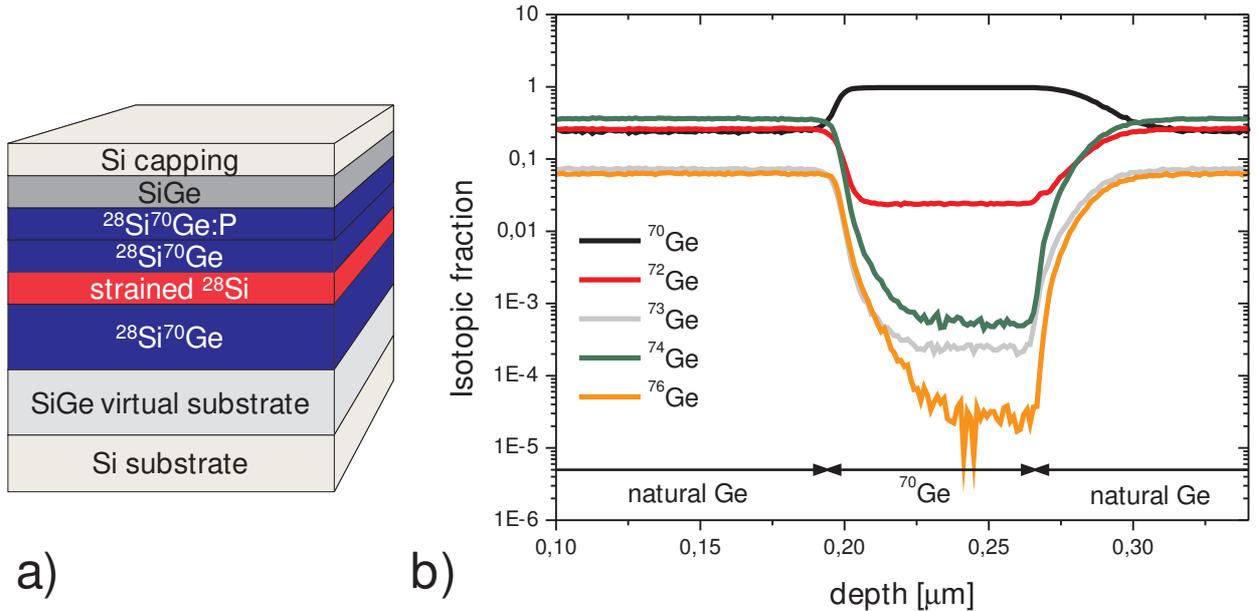}
\caption{a) Layer structure of an isotopically engineered Si/SiGe heterostructure. The 2DES forms in the strained $\mathrm{^{28}Si}$ layer in a nuclear spin free environment. b) Isotopic abundances of all five naturally occuring Ge isotopes in a MBE grown layer structure in which a thick layer of $\mathrm{^{70}Ge}$ is embedded into natural Ge. In the $\mathrm{^{70}Ge}$ layer, the only nuclear spin carrying isotope $\mathrm{^{73}Ge}$ decreases from approx. 8\% to approx. $\mathrm{2.5 \cdot 10^{18}~cm^{-3}}$.}
\label{Iso}
\end{figure}

Fig.~\ref{Iso}~a) shows the typical layout of our isotopically engineered Si/SiGe heterostructures. A SiGe virtual substrate of natural isotopic composition is first deposited onto a (100) oriented Si substrate. The virtual substrate is realized by increasing the Ge content linearly by $\mathrm{8\%/{\mu}m}$ until the desired Ge content is reached. The graded layers are deposited at a substrate temperature of $\mathrm{T_s~=~575~^{\circ}C}$.The virtual substrate is fully relaxed within the experimental error of $\mathrm{10\%}$. It typically displays a density of threading dislocations of about $\mathrm{1 \cdot 10^6~cm^{-2}}$. The active part is then realized in the same fabrication process but from isotopically purified material. It is composed of a $\mathrm{15~nm}$ thick, fully strained and dislocation free $\mathrm{^{28}Si}$ layer embedded into a $\mathrm{^{28}Si^{70}Ge}$ cladding. The lower part of the cladding separates the nuclear spins of the virtual SiGe substrate from the 2DES. The upper part of the cladding contains a modulation doping realized with a $\mathrm{15~nm~^{28}Si^{70}Ge}$ spacer layer and a $\mathrm{15~nm}$ volume doped $\mathrm{{^{28}Si^{70}Ge:P}}$ layer. The P concentration amounts to $\mathrm{1 \cdot 10^{18}~cm^{-3}}$. The growth temperature $\mathrm{T_s}$ was lowered before depositing the $\mathrm{{^{28}Si^{70}Ge:P}}$ and all subsequent layers to prevent P segregation to the surface, which could result in a possible lowering of the Schottky barrier to the gate metal. Finally, the active part of the heterostructure is capped with $\mathrm{45~nm}$ natural SiGe and protected against oxidation by a $\mathrm{10~nm}$ thick Si layer on top.

The residual contamination with the nuclear spin carrying isotopes $\mathrm{^{29}Si}$ and $\mathrm{^{73}Ge}$ of both enriched single crystalline source materials, $\mathrm{^{28}Si}$ and $\mathrm{^{70}Ge}$, respectively, is below 0.1\%. The abundance of all five naturally occurring Ge isotopes in a MBE grown layer structure has been determined by high resolution secondary ion mass spectrometry (SIMS) and is exemplarily shown in Fig.~\ref{Iso}~b). To meet the requirements for high concentration resolution SIMS, a specially designed, MBE grown trilayer consisting of a thick $\mathrm{^{70}Ge}$ layer sandwiched between two layers of Ge of natural isotopic composition was analyzed. In this trilayer, the abundance of the nuclear spin carrying isotope $\mathrm{^{73}Ge}$ decreases by almost 3 orders of magnitude from approx. 8\% in the natural Ge to below 0.025\%, i.e. $\mathrm{2.5 \cdot 10^{18}~cm^{-3}}$, in the $\mathrm{^{70}Ge}$ layer. An equivalent suppression of the number of nuclear spins has also been verified for MBE grown layers involving the $\mathrm{^{28}Si}$ source.

Electrical characterization of the 2DES has been done in a $\mathrm{^{3}He}$ cryostat at temperatures down to $\mathrm{320~mK}$ and magnetic fields up to $\mathrm{10~T}$ in Hall bar geometry.  $\mathrm{20~{\mu}m}$ wide Hall bars are defined photo-lithographically and wet-etched in a solution of diluted hydroflouric acid and concentrated nitric acid. The Ohmic contacts are formed by an {Au/Sb/Au} $\mathrm{20/300/1100}$~{\AA} trilayer which is annealed at $\mathrm{T~=~450~^{\circ}C}$ in an inert $\mathrm{N_2}$ atmosphere for $\mathrm{180~s}$ and covered with a TiAu bilayer. Longitudinal ($\mathrm{U_{xx}}$) and transversal ($\mathrm{U_{xy}}$) voltage drops were acquired at the same time using two lock-in amplifiers.

\begin{figure}[b]
\includegraphics[width=0.8\columnwidth]{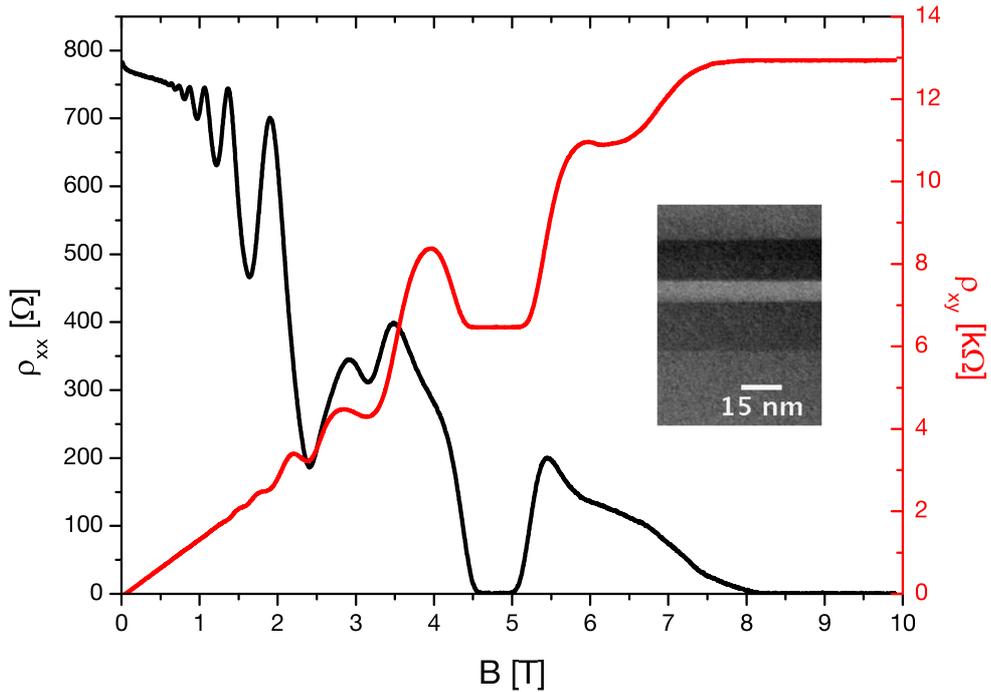}
\caption{Longitudinal (black) and transversal (red) Hall resistance for the isotopically engineered 2DES measured at $\mathrm{T~=~340~mK}$. Inset: TEM micrograph of the active part of the heterostructure. The strained $\mathrm{^{28}Si}$ layer appears bright. In the SiGe layers, stronger contrasts correspond to a higher Ge content.}
\label{Transport}
\end{figure}

Fig.~\ref{Transport} shows a typical result obtained during a magnetic-field sweep from $\mathrm{B~=~0~T}$ to $\mathrm{B~=~10~T}$ at $\mathrm{T~=~340~mK}$ after illumination of the sample until saturation of the charge carrier density. Relevant parameters deduced from the measurement prove the high quality of the 2DES. The 2D sheet carrier density obtained from the low field slope of the Hall resistance $\mathrm{\rho_{xy}(B)}$ corresponds very well to the density obtained by a Fourier transform of the Shubnikov-de Haas oscillations in $\mathrm{\rho_{xx}}$ and amounts to $\mathrm{4.6 \cdot 10^{11}~cm^{-2}}$. Shubnikov-de Haas oscillations in $\mathrm{\rho_{xx}}$ start to develop at B-fields as low as $\mathrm{0.6~T}$ and show the 4-fold periodicity characteristic for 2DES in Si/SiGe, originating from both, the valley- and spin-degeneracy. The highest filling factors observed are 32 in $\mathrm{\rho_{xx}}$ and 12 in $\mathrm{\rho_{xy}}$. Spin split levels can be resolved for filling factors lower than 10. The shoulders in $\mathrm{\rho_{xx}}$ between filling factors 6 and 4 as well as 4 and 2 indicate valley splitting. In the high B-field regime, two very well defined quantum Hall effect plateaus with corresponding minima of the Shubnikov-de Haas oscillations for filling factors four and two are visible. Our structures show no sign for any parallel conduction which might for example arise in the dopant supply layer. This absence of parallel conduction has additionally been confirmed by Hall mobility spectrum analysis measurements \cite{Myr2008}. The remarkable Hall resistance overshoot visible before filling factors 3, 4, 6 and 8 is generally observed in our Si/SiGe heterostructure 2DES with narrow Hall bars and is not induced by the use of isotopically enriched source material. It has also been observed by other groups \cite{Gri2000,Shl2005} and not only for the Si/SiGe material system \cite{RIC1992,Poi2004}, but no general picture of its origin has emerged yet. A systematic variation of the Hall bar geometry and experimental parameters we have carried out on Si/SiGe 2DES both with natural and isotopically engineered compositions points towards a phenomenon taking place at the Hall bar edge. The overshoot should thus not impede the realisation of few to one electron devices in this material. The systematic study will be published elsewhere.

The zero-field Hall mobility of the sample shown in Fig.~\ref{Transport} is $\mathrm{18000~cm^2/Vs}$. A scattering time analysis  \cite{BOC1990,TOB1992,TOB1992a} performed to determine the main scattering source, suggests that long range remote impurity scattering is the main mobility limiting scattering mechanism. Nevertheless some short range scattering contributions are also observed which are absent in our Si/SiGe heterostructures of natural isotopic composition. The probable origins are found through a structural analysis. Cross-sectional transmission electron microscopy (TEM) analysis as shown in the inset of Fig.~\ref{Transport} for example reveals a material contrast in the active part. Due to the imaging conditions set, strong contrasts correspond to a higher Ge content. This indicates a slight mismatch in the Ge content between the $\mathrm{{^{28}Si^{70}Ge}}$ layers in the active area and the virtual SiGe substrate as well as the SiGe capping respectively. The resulting lattice mismatch induces potential fluctuations leading to the observed contributions of short range scattering. A suppression of this mismatch via the implementation of a more sophisticated flux control of the four different sources involved in the fabrication, Si, Ge, $\mathrm{^{28}Si}$ and $\mathrm{^{70}Ge}$, should eliminate the short range scattering contribution and in turn lead to higher mobilities in future structures.

\begin{figure}[t]
\includegraphics[width=0.8\columnwidth]{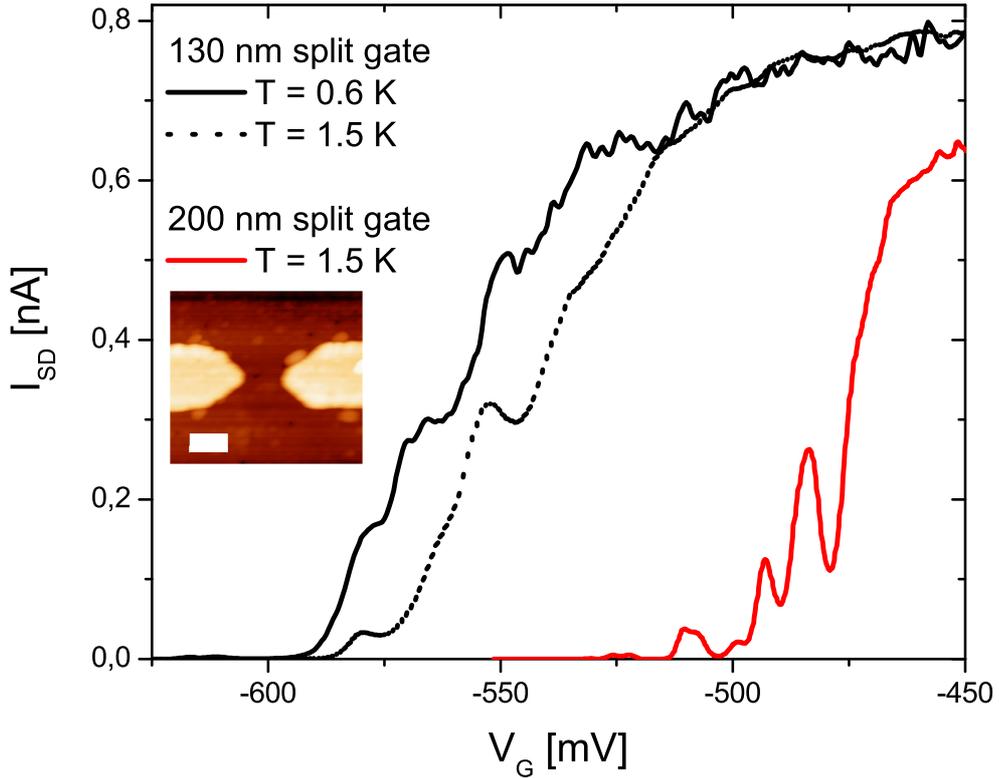}
\caption{Transport through two different split-gates defined on the same isotopically engineered Si/SiGe heterostructure. Measurements have been done after different cool-downs of the sample and in two different measurement setups. Inset: AFM micrograph of a nano-constriction. The scale-bar corresponds to $\mathrm{130~nm}$.}
\label{gate}
\end{figure}

Electron beam written symmetric split gates with tip-to-tip distances ranging from $\mathrm{130~nm}$ to $\mathrm{200~nm}$, as for example shown in the AFM micrograph in the inset of Fig.~\ref{gate}, have been realized on nuclear spin free 2DES with palladium (Pd). The Pd gates proved to be very stable and reliable over time even after several thermal cycles of the sample between room temperature and cryogenic temperatures. For all samples, the leakage current of the Pd gates was below the detection limit of $\mathrm{2~pA}$ of our measurement setups for the whole range of operation of the gates. Remarkably, the Pd gates deplete the 2DEG underneath already at zero applied bias (not shown). Fig.~\ref{gate} shows results obtained for electrical transport along a narrow conducting channel defined by the narrowest and the widest, $\mathrm{130~nm}$ and $\mathrm{200~nm}$ wide, nano-constrictions fabricated on one Hall-bar. The measurements have been taken on different days. Complete pinch-off of the narrow conducting channels is achieved for gate voltages as low as $\mathrm{-0.6~V}$ and $\mathrm{-0.5~V}$ respectively, opening the possibility to design a Schottky top-gated quantum dot. We attribute the conductance steps and peaks to potential fluctuations and thereby induced Coulomb blockade in the vicinity of the narrow conducting channel \cite{Tob1995a}.

In summary, employing $\mathrm{^{28}Si}$ and $\mathrm{^{70}Ge}$ source material in MBE, we have realized a 2DES in a nuclear spin free environment in strained $\mathrm{^{28}Si}$. Low-temperature magneto-transport measurements prove the high quality of the 2DES with well defined quantum Hall effect plateaus. Reliable control of the 2DES and pinch-off a narrow conducting channels has been demonstrated using split-gates. Our devices represent a promising basis to study the impact of the absence of nuclear spins on the decoherence of single electron spins by means of electrostatically top-gate defined quantum dots.

The authors gratefully acknowledge H. Cerva at Siemens AG Corporate Technology for access to electron microscopy facilities and financial support by the Deutsche Forschungsgemeinschaft via SFB631 and the Excellence Cluster Nanosystems Initiative Munich (NIM). The work at Keio was supported in part by MEXT program No. 18001002, by Special Coordination Funds for Promoting Science and Technology, and by Grant-in-Aid for the Global Center of Excellence. Work at the LBNL was supported in part by US NSF Grant Nos. DMR-0405472 and the U.S. DOE under Contract No. DE-AC02-05CH11231.

\removelastskip

\end{document}